\documentclass[aps,twocolumn]{revtex4}

\usepackage{enumerate}

\usepackage{amsmath}

\begin{document}

\title{Comment on "Time-dependent   entropy  of simple quantum model systems"}
\author{Piotr Garbaczewski\thanks{electronic address: p.garbaczewski@if.uz.zgora.pl}}
\affiliation{Institute of Physics,  University  of Zielona
G\'{o}ra, 65-516 Zielona G\'{o}ra, Poland}

\begin{abstract}
In the above mentioned paper by J. Dunkel and S. A. Trigger [Phys. Rev.  {\bf A 71}, 052102, (2005)]
a hypothesis has been pursued  that  the
loss of information associated with the quantum  evolution of pure states, quantified in terms of
an  increase in time  of so-called Leipnik's joint entropy, could be a rather  general
 property shared by many quantum systems.  This behavior has been confirmed  for the
 unconfined model systems and properly tuned initial data (maximally classical states).
  We provide two particular examples which indicate a complexity of the  quantum evolution.
  In the presence of a confining (harmonic) potential Leipnik's entropy may be non-increasing
for maximally classical initial data.  Another choice of initial data  implies  periodicity
 in time of the Leipnik entropy.
\end{abstract}
\maketitle

 \vskip0.3cm

The von Neumann entropy is  insensitive to the unitary quantum dynamics and in addition
 vanishes  for   pure states.
Hence, from the modern  quantum   information-theory point of view   pure quantum  states
(including elaborate  text-book discussions of the  wave-packet dynamics)  appear  to be  useless.
However,  there are available  other  information-theoretic entropy tools  which   quantify both
the "information content" (here, complementarily  interpreted as the "uncertainty content")
 of  quantum wave-packets and give account of their  Schr\"{o}dinger picture dynamics.

A straightforward generalization of the Shannon entropy for Born postulate-inferred  continuous
probability distributions, named the Leipnik entropy,  has been used in Refs. \cite{dunkel,trigger}
to quantify the loss of information associated with temporally evolving pure quantum states.
For simple  model system examples considered by \cite{dunkel} ,  initial time $t=0$ states
were chosen to minimize both the Heisenberg uncertainty relation and the joint entropy. They
were  named MACS (maximal classical states).

For unconfined model systems investigated in \cite{dunkel}, it was found that:  "A quantum system
that has been in a MACS at time $t=0$ inevitably evolves into a non-MACS at times $t>0$. This
intrinsic property of quantum systems is e.g. reflected by a monotonous increase of the joint entropy."
This particular observation has led the Authors  to conclude that "most likely,
this quantum trend also manifests itself for other types of initial wave packets and external
 potentials, as well  as in many-particle systems".

In the present Comment we  wish to indicate  that, in  the presence of the confining (harmonic)
 potential,    the MACS initial data  may  preserve the MACS property in time, and thence the
   Leipnik  entropy  may be  conserved in the course of the quantum  evolution. Another initial
   data choice implies that the Leipnik entropy is  periodic in  time.

Given  an $L^2(R)$-normalized function  $\psi (x)$ we denote $({\cal{F}}\psi )(p)$ its Fourier
transform. The corresponding probability densities follow:
 $\rho (x) = |\psi (x)|^2$ and $\tilde{\rho }(p) = |({\cal{F}}\psi )(p)|^2$.\\
We introduce the related position and momentum  information  (differential) entropies:
${\cal{S}}(\rho )\doteq S_q = - \int \rho (x) \ln \rho (x)  dx $    and   ${\cal{S}}(\tilde{\rho })\doteq
S_p= - \int \tilde{\rho }(p) \ln \tilde{\rho }(p)  dp $, where  ${\cal{S}}$ denotes the  Shannon
 entropy for a continuous probability  distribution, also named differential entropy.

We assume both entropies to take finite values. Then,
there holds the familiar   entropic uncertainty relation \cite{mycielski}:
\begin{equation}
S_q + S_p \geq  (1 + \ln \pi ) \, .
\end{equation}
Up to an irrelevant  additive constant (we shall pass to natural units with $\hbar \equiv 1$),
the left-hand-side of the entropic  inequality coincides
with the Leipnik joint entropy  $S_J$  discussed in ref. \cite{dunkel}.

 In the above, no explicit time-dependence has been  indicated, but all derivations go through with
any wave-packet  solution $\psi (x,t)$ of the Schr\"{o}dinger equation. The  induced dynamics of
 probability densities
may imply the  time-evolution of  entropies: $S_q(t), S_p(t), S_J(t)$.

If, following  conventions we define
the squared  standard  deviation   value  for an observable $A$ in a pure
state $\psi $ as
$(\Delta A)^2 = (\psi , [A - \langle A\rangle ]^2 \psi )$ with $\langle A
\rangle  = (\psi , A\psi)$, then for the  position $X$ and momentum $P$ operators
we have the following version of the  entropic uncertainty relation  (here expressed
through so-called entropy powers, see e.g.   \cite{petz}, $\hbar \equiv 1$):
\begin{equation}
\Delta X \cdot \Delta P \geq  {\frac{1}{2\pi e}} \,
 \exp[{\cal{S}}(\rho )  + {\cal{S}}(\tilde{\rho })] \geq {\frac{1}{2}}
\end{equation}
which is  an alternative version of the entropic uncertainty relation.

An important property of the Shannon entropy  ${\cal{S}}(\rho ) $ is that for any   general
 probability distribution $\rho (x)$ with a  fixed variance $\sigma $
we would have ${\cal{S}}(\rho )\leq  {\frac{1}2}  \ln (2\pi e \sigma ^2)$.
${\cal{S}}(\rho )$ becomes maximized  in the set of such densities  if and only if $\rho $
is a Gaussian  with  variance $\sigma$.  For Gaussian densities  $ (2\pi e )\Delta X \cdot \Delta P =
 \exp[{\cal{S}}(\rho )  + {\cal{S}}(\tilde{\rho })]$ holds true, but the     minimum $1/2$
  on the right-hand-side of Eq.~(2),   is  not necessarily  reached.

The familiar (Schr\"{o}dinger's)   coherent state of the  harmonic oscillator has
the property of transforming  \it  all  \rm  inequalities of Eq. (2) into an identity, and   preserves coherence
 for all times. Therefore, we have in hands  a  non-stationary solution of the Schr\"{o}dinger
 equation for which  $S_J$ is manifestly  time-independent, although being the MACS in the
 terminology of Ref.  \cite{dunkel}. The MACS property is left intact by the
 Schr\"{o}dinger time evolution.

 Assuming  for simplicity $\hbar =m=\omega =1$ we have:
\begin{equation}
\rho (x,t) = \pi ^{-1/2} \exp [-(x- q(t))^2]
\end{equation}
 where  $q(t)=q_0 \cos t + p_0 \sin t$.  In the present case, \cite{majernik}:
\begin{equation}
  S_q(t) = S_p(t) =  (1/2) (1+\ln \pi ) \, .
 \end{equation}
The monotonous increase  of the Leipnik entropy is   \it not \rm  an  inevitable consequence of the
 MACS initial data.

The previously mentioned  "quantum trend"  as well   may not arise  for other types
of initial wave packets. We shall give a non-MACS example  where the indeterminacy relation is saturated,
but the Leipnik entropy is not at its  extremum.

Let us  consider  the squeezed  wave function of the  harmonic oscillator, \cite{majernik}.
We work with  the re-scaled units $\hbar =\omega =m=1$.
 The solution of the Schr\"{o}dinger
equation $i\partial _t\psi = -(1/2)\Delta \psi  + (x^2/2)\psi$  with the initial data $\psi (x,0) = (\gamma ^2
\pi )^{-1/4} \exp(-x^2/2\gamma ^2)$  and $\gamma \in (0,\infty )$,   gives rise to  the time-dependent
probability density :
\begin{equation}
\rho (x,t) = {\frac{1}{(2\pi )^{1/2}\sigma(t)}} \exp \left( - {\frac{x^2}{2\sigma ^2(t)}} \right)
\end{equation}
where
\begin{equation}
2\sigma ^2(t) = {\frac{1}{\gamma ^2}} \sin ^2t + \gamma ^2 \cos^2t \, .
\end{equation}

The position  entropy  reads  $S_q= (1/2) \ln [2\pi e \sigma ^2(t)]$.
The momentum entropy $S_p$  has the same functional form as $S_q$, except for the replacement
 of $\sigma ^2(t)$ by
${\tilde{\sigma }}^2(t) = \gamma ^2 \sin ^2t + (1/\gamma ^2) \cos ^2t$ (that is special to the harmonic
oscillator case).

As a consequence, the Leipnik entropy  $S_J=S_q+ S_p$ is a periodic
 function of time with a period $\pi /2$
and oscillates between $1 + \ln \pi $ and $1+\ln \pi + (1/2)\ln [(1/4)(\gamma ^4 + \gamma ^{-4} + 2)]$,
\cite{majernik}.
A similar periodic behavior has been found for  prototype Schr\"{o}dinger cat states,  initially  modelled
as a superposition of two Schr\"{o}dinger coherent states with the same amplitude but opposite phases.

It is useful to mention that  the arithmetic sum of position and momentum entropies has received an ample
attention in the literature investigating complex atoms, \cite{panos} while the  general  issue of the
  Shannon entropy dynamics  has been addressed in  \cite{gar}.

Let us add  some general comments about the uses of time-dependent entropies in quantum theory.

Among numerous  manifestations of the concept of
entropy in physics and mathematics, information-theory based
entropy methods were devised to investigate the large time
behavior of solutions for various (mostly dissipative)  partial
differential equations.
Shannon, Kullback and von Neumann entropies are typical information
 theory   tools  which  were  designed  to quantify  the "information
content" and possibly "information loss" for  systems in a specified state.

  For quantum systems, the von Neumann entropy vanishes on
 pure states, hence   one presumes to have   a "complete information" about
the state.  On the other  hand, for pure states
the differential entropy  gives access
to another "information  level", associated with a probability
distribution inferred from  a  $L^2(R^n)$  wave packet. It is
perfectly suited   to give account of the Schr\"{o}dinger picture
dynamics and this property extends   to  the  Leipnik entropy.

 Since, in  physical sciences,  entropy is  typically regarded
 as  a measure of
the degree of randomness and the  tendency (trends)  of physical
systems to become less and less "organized", it is   quite  natural
to think of entropy as about the measure of uncertainty.
In view of the profound role played by the Shannon entropy  in the
formulation of entropic indeterminacy relations,   the term
"information", in the present context should be  used  in the technical sense, meaning
the inverse of "uncertainty".

We may attribute  a concrete  meaning to the term "organization" in the quantum  wave packet context.
Namely, Shannon and Leipnik entropies  quantify  the degree of  the
probability distribution   "complexity", \cite{panos},  and  "(de)localization",  \cite{gar},
for stationary and non-stationary Schr\"{o}dinger wave packets.

{\bf  Acknowledgement:} This note  has been supported by the Polish Ministry of Scientific Research and
Information Technology under the  (solicited) grant No PBZ-MIN-008/P03/2003.


\begin{thebibliography}{99}
\bibitem{dunkel} J. Dunkel and S. A. Trigger, Phys. Rev. {\bf A 71}, 052102 (2005)
\bibitem{trigger} S. A. Trigger, Bull. Lebedev. Phys. Inst. {\bf 9}, 44 (2004)
\bibitem{majernik} V. Majernik and T. Opatrn\'{y}, J. Phys. A: Math. Gen. {\bf 29}, 2187 (1996)
\bibitem{mycielski} I. Bia{\l}ynicki-Birula and J.  Mycielski,  Commun. Math. Phys.  {\bf 44}, 129 (1975)
\bibitem{petz} Ohya, M. and Petz, D., {\it Quantum Entropy and Its use},  Springer-Verlag, Berlin, 1993
\bibitem{panos} K. Ch. Chatzisavvas , Ch. C. Moustakidis and C. P. Panos, Information entropy,
information distances and complexity of atoms, arXiv:quant-ph/0507039, (2005)
\bibitem{gar} P. Garbaczewski, Differential entropy  and dynamics of uncertainty, arXiv:quant-ph/0408192,
(2004)
\end{thebibliography}
\end{document}